# High-pressure, temperature elasticity of Fe- and Al-bearing MgSiO$_3$: implications for the Earth's lower mantle

Shuai Zhang[a,*], Sanne Cottaar[b], Tao Liu[c], Stephen Stackhouse[c], Burkhard Militzer[a,d]

[a]*Department of Earth and Planetary Science,*

*University of California, Berkeley, California 94720, USA*

[b]*Department of Earth Sciences, University of Cambridge, Cambridge, UK*

[c]*School of Earth and Environment, University of Leeds, Leeds LS2 9JT, UK*

[d]*Department of Astronomy, University of California, Berkeley, California 94720, USA*

## Abstract

Fe and Al are two of the most important rock-forming elements other than Mg, Si, and O. Their presence in the lower mantle's most abundant minerals, MgSiO$_3$ bridgmanite, MgSiO$_3$ post-perovskite and MgO periclase, alters their elastic properties. However, knowledge on the thermoelasticity of Fe- and Al-bearing MgSiO$_3$ bridgmanite, and post-perovskite is scarce. In this study, we perform *ab initio* molecular dynamics to calculate the elastic and seismic properties of pure, Fe$^{3+}$- and Fe$^{2+}$-, and Al$^{3+}$-bearing MgSiO$_3$ perovskite and post-perovskite, over a wide range of pressures, temperatures, and Fe/Al compositions. Our results show that a mineral assemblage resembling pyrolite fits a 1D seismological model well, down to, at least, a few hundred kilometers above the core-mantle boundary, i.e. the top of the D″ region. In D″, a similar composition is still an excellent fit to the average velocities and fairly approximate to the density. We also implement polycrystal plasticity with a geodynamic model to predict resulting seismic anisotropy, and find post-perovskite with predominant (001) slip across all compositions agrees best with seismic observations in the D″.

\* Corresponding author.

*E-mail address*: shuai.zhang01@berkeley.edu (S. Zhang)

# 1. Introduction

Fe-bearing perovskite (Pv, a.k.a. bridgmanite) and post-perovskite (pPv) are the most abundant minerals in the Earth's lower mantle. A good knowledge on their elastic properties under relevant high-pressure (P) and temperature (T) conditions is of fundamental importance for understanding the structure and dynamics of the Earth's interior. Experiments have suggested that Pv hosts ~10 mol.% iron and a slightly lower amount of Al in the pyrolitic model composition, expected from the top to, at least, the mid-lower mantle (Irifune et al., 2010). Despite of this, it is still debated how much iron remains in Pv, how it is distributed between Pv and pPv, and whether $Fe^{2+}$ disproportionates to form $Fe^{3+}$+Fe, or partitions into magnesiowüstite or melt, in the lowermost mantle (e.g., Mao et al., 2004; Irifune et al., 2010). Currently, there is scarce experimental data available on the elastic properties of these systems (Murakami et al., 2012), and the uncertainty could be huge, due to the interplay between the above factors and technical issues relating to the extreme conditions. This necessitates theoretical methods, which are not limited by the same constraints as experimental and have been proven to be a powerful tool in determining the elastic properties of minerals at lower mantle conditions.

There have been several reports on the elastic properties of these Pv and pPv systems using first-principle calculations. However, these have mainly focused on pure $MgSiO_3$ for high-T conditions (Oganov et al., 2001; Wentzcovitch et al., 2004, 2006; Stackhouse et al., 2005; Stackhouse and Brodholt, 2007; Zhang et al., 2013), or Fe/Al-bearing systems at 0 K (Kiefer et al., 2002; Caracas and Cohen, 2005; Li et al., 2005a; Stackhouse et al., 2005, 2006, 2007; Tsuchiya and Tsuchiya, 2006), because of the high computational cost and the complex structural and electronic properties of Fe/Al-bearing systems at high P and T.

Examples on this complexity include the pressure-induced high-spin (HS) to low-spin (LS) transition in iron (Badro et al., 2004) and challenges arising from the large number of mechanisms that iron can enter the Pv and pPv structures. Ferrous iron substitutes for Mg in the pseudo-dodecahedral A (Mg-) site, but for ferric iron charge balance requires the coupled substitution of Fe-Al or Fe-Fe pairs (Zhang and Oganov, 2006a), with one cation at the A site and the other at the octahedral B (Si-) site. Incorporation of ferric iron via the oxygen vacancy mechanism is not expected in Pv and pPv at lower mantle conditions (Brodholt, 2000).

Ferrous iron, in the A site, is expected to remain in the HS state, in Pv and pPv, at all mantle conditions (Hsu et al., 2011a; Yu et al., 2012). In contrast, ferric iron undergoes a spin transition at lower-mantle pressures, as the crystal field splitting surpasses the spin pairing energy, which breaks the Hund's rule that predicts all five $d$ electrons in $Fe^{3+}$ have the same spin (HS state) and stabilizes the LS state (3 electrons spin up, 2 down). The spin transition of ferric iron is expected to start at about 75 GPa at the A site, and at much lower pressures at the B site (Zhang and Oganov, 2006b; Li et al., 2005a; Caracas, 2010; Hsu et al., 2011b, 2012).

In recent years, there have been several studies on the effects of ferrous and ferric iron on the thermoelastic properties of Pv and pPv (Metsue and Tsuchiya, 2011; Tsuchiya and Wang, 2013; Shukla et al., 2015). To date, agreement with seismological models of the lower mantle has not been reached. One potential reason is that the influence of $Al^{3+}$ has not yet been included, and another could be related to the possible inadequacy of using the quasi-harmonic approximation (QHA) in computing thermoelasticity at high-P and T. There has been one study that used density functional molecular dynamics (DFT-MD) to calculate the properties of Pv and pPv enriched with ferrous iron (Muir and Brodholt, 2015), but this only reports values at one pressure.

In this letter, we apply DFT-MD to determine the elastic properties of ferrous and ferric iron-bearing Pv and pPv over a wide range of pressure and temperature conditions relevant to the Earth's lower mantle. In the case of ferric iron, Fe-Al and Fe-Fe substitution are considered. DFT-MD naturally accounts the anharmonicity of lattice dynamics, which is important for describing the elastic behavior of minerals at high T (above the Debye temperature) and has been used in several previous studies (Stackhouse et al., 2005; Zhang et al., 2013). In the case of ferrous iron only the HS state is considered, whereas for ferric iron different Fe-Fe or Fe-Al compositions in various possible spin states are considered. High-T lattice parameters are obtained by performing DFT-MD simulations in the NPT ensemble, which is followed by NVT calculations on strained structures to calculate the elastic coefficients and other elastic properties. Based on these calculations, we derive the seismic properties of a mineral assemblage to compare with a 1D seismological model, with the aim of constraining the mineral composition of the lower mantle. We further exemplify shear wave anisotropic properties resulting from deformation within a deep subducting slab and discuss the implications for seismic observations of shear wave anisotropy.

## 2. Theory

*2.1. Computational methods*

Our calculations are performed using the Vienna *ab initio* simulation package (VASP) (Kress and Furthmüller, 1996). Projector augmented wave (PAW) pseudopotentials (Blöchl, 1994) and the local density approximation (LDA)-based (Ceperley and Alder, 1980) exchange-correlation functional are used. It has been shown that in some cases, standard DFT functionals can fail to predict the correct electronic properties of iron-bearing minerals, and inclusion of a Hubbard U term leads to improved predictions of their electronic, structural and elastic properties (Stackhouse et al., 2010). However, a recent study suggests that provided standard DFT functionals produce the correct insulating ground states and correct orbital occupancy, the structural and elastic properties predicted with them should be similar to those with a U term (Hsu et al., 2011a). We confirmed this conclusion by comparing the results of calculations with and without a U term, for a select number of iron-bearing Pv structures (see Table S4 in the supplementary material), and performed the remainder without a U term.

The supercell sizes 2×2×1 (80-atom) for Pv, and 3×1×1 (60-atom) or 4×1×1 (80-atom) for pPv are used. We employ pseudopotentials with core radii equaling 2.2, 1.9, 2.0, 1.9, and 1.52 Bohr for Fe, Al, Mg, Si, and O, respectively. In static calculations, the plane wave energy cutoff is set to 800 eV, and Monkhorst-Pack $k$-mesh (Monkhorst and Pack, 1976) of 2×2×4 (for Pv supercell) or 6×6×6 (for pPv supercell) are chosen for Brillouin zone sampling. All structures are relaxed until all forces are smaller than $10^{-4}$ eV/Å, based on which the enthalpy is evaluated. In MD simulations, we follow previous studies (Stackhouse and Brodholt, 2007; Zhang et al., 2013) and use time step of 1 fs, energy cutoff of 500 eV and only the Γ point to sample the Brillouin zone. Tests show that using these settings the elastic properties are converged to within statistical error. The lattice constants are obtained by averaging over NPT trajectories. These are then used in the NVT simulations. For calculating the elastic constants from linear stress-strain relations, three axial and one triclinic strain are applied to the cell. During NPT, we fix the cell to be orthorhombic to avoid unnecessary fluctuations. The same averaging scheme is applied to NVT trajectories to retrieve stress components and energies. The NPT and NVT trajectories are typically over 3 ps and 5 ps, respectively. The convergence is shown in the supplementary material.

For ferrous iron we include four $Fe^{2+}$ in 80-atom unit cells of Pv and pPv to construct $(Mg_{1-x}Fe_x)SiO_3$ structures with an atomic percent $x$ equal to 25%. For ferric iron we include one or two $Fe^{3+}$-$M^{3+}$ (M representing Fe or Al) pairs in $MgSiO_3$ by charge-coupled substitution $Fe_{Mg}+M_{Si}$ in order to construct $(Mg_{1-x}Fe_x)(Si_{1-x}M_x)O_3$ structures with atomic percent $x$ equal to 6.25% or 12.5% for Pv, and 8.33% or 16.7% for pPv. Such iron or aluminum compositions are relevant to the lower mantle (Irifune et al., 2010). For $Fe^{2+}$ in these models, we consider only the HS (spin momentum S=4/2) ferromagnetic state, whereas for $Fe^{3+}$ we consider both the HS (spin momentum S=5/2) and LS (S=1/2) states. Intermediate-spin states are not considered because they have been found to be energetically disfavored in previous studies (Hsu et al., 2011b). For systems with multiple ferric iron cations, we take into account all possible spin configurations, including pure HS, pure LS, or HS-LS mixtures in the ferromagnetic (FM) or antiferromagnetic (AFM) states. Our results confirm conclusions reached in several previous DFT calculations (Zhang and Oganov, 2006b) on the site-dependence and values of the transition pressure, i.e., A-site $Fe^{3+}$ remains HS until very high (>120 GPa) pressure and B-site $Fe^{3+}$ is likely to be in the LS state in the lower mantle. This enables us to disregard unlikely configurations, for example, B-site $Fe^{3+}$ in HS at 100 GPa. In spite of this, there are still an ample amount of configurations to consider.

For ferric iron-bearing structures, we first perform a structural relaxation at 0 K by fixing the spin state of each iron in the system. For the different compositions, we compare the enthalpies of the different spin states at each pressure, in order to determine the most stable configurations. The magnetic entropy is the same for $Fe^{3+}$ at either A or B site whether it is HS or LS. If we assume the vibrational entropy of the different spin configurations to be similar, then the enthalpies provide a good approximation to the most stable spin states at high temperature without having to calculate the Gibbs free energy. We limit our studies of high temperature elasticity mainly to these configurations. In spite of this, because there does not exist a perfect way of describing the correlation effects of $3d$ electrons of iron and temperature could potentially stabilize the HS state, at high temperature we selectively carry out additional calculations on configurations other than the most stable ones at 0 K, in order to get a more complete picture on the effect of different magnetic states on the elastic properties.

*2.2. Calculation of elastic, seismic, and thermodynamic properties*

The elastic constants are determined assuming the linear stress-strain relation $\sigma_\alpha \propto C_{\alpha\beta}\epsilon_\beta$, although higher-order terms become more significant for larger strains. In our calculations, we choose an optimized strain of magnitude $\epsilon = \pm 0.5\%$ (Militzer et al., 2011). The orthorhombic symmetry of Pv and pPv allow us to obtain the nine independent elastic constants by applying three axial strains and one triclinic strain (Stackhouse and Brodholt, 2007). The polycrystalline bulk and shear moduli, $K$ and $G$, are determined consequently using the Voigt averaging scheme. A comparison with the Voigt-Reuss-Hill (VRH) scheme is shown in the supplementary Table S5.

The direction-dependence of the phase velocities is governed by the Christoffel equation $[\rho v^2 \delta_{ik} - C_{ijkl} n_j n_l][u_k] = 0$, where $C_{ijkl}$ is the elastic tensor where the indexes $ij$ and $kl$ correspond to the above $\alpha$ and $\beta$ via the Voigt notation: $\{11; 22; 33; 23; 13; 12\} \rightarrow \{1; 2; 3; 4; 5; 6\}$, $\delta_{ik}$ is the Kronecker delta and $\{n_j\}$ denotes the wave propagation direction. The solution to the Christoffel equation yields three eigenvalues that are the phase velocities of the three wave modes, P, SV and SH, that can exist in an anisotropic solid. Because of the high velocity, the transmission of seismic waves in the lower mantle is primarily an adiabatic process rather than an isothermal one. Therefore, we add an adiabatic correction according to (Wallace, 1972)

$$C_{\alpha\beta}^{S} = C_{\alpha\beta}^{T} + \frac{TV}{C_{\eta}} b_{\alpha} b_{\beta}, \tag{1}$$

where $b_{\alpha} = \left(\frac{\partial \sigma_{\alpha}}{\partial T}\right)_{\eta}$, $C_{\eta}$ is fixed-configuration heat capacity approximated by $C_V = \left(\frac{\partial E}{\partial T}\right)_V$, which can be obtained by two additional simulations for each structure with fixed volume and temperatures changed by ±100 K. The Grüneisen parameter and the thermal expansion coefficient can also be estimated from these simulations via $\gamma = V\left(\frac{\partial P}{\partial E}\right)_V$ and $\alpha = \frac{\gamma \rho C_V}{K}$, respectively.

In order to compare the results with experimental data, one has to apply a pressure correction, because of the well-known fact that LDA underestimates lattice constants and thus the pressure. One simple way is to calculate at experimentally determined ambient-pressure volume in the static condition, and then apply the pressure difference as a constant shift (Oganov et al., 2001) $\Delta P$ to all computed pressures regardless of $P$, $T$, Fe/Al compositions and the mineral phases. It is worthwhile to note that this is by no means a perfect solution, but works reasonably well (Stackhouse and Brodholt, 2007; Militzer et al., 2011; Zhang et al., 2013). With a careful estimation on the ambient-pressure volume in the static condition (see supplementary material), we determined $\Delta P = 4.0 \pm 0.8$ GPa.

### 3. Results and discussion

*3.1. Phase relations*

The most stable spin configurations at 0 K for the ferric iron systems are summarized in Table 1. In $(Mg_{1-x}Fe_x)(Si_{1-x}Fe_x)O_3$ Pv, $Fe^{3+}$ at the A-site undergoes a spin transition below 50 GPa; when including $Al^{3+}$, the spin transition pressure increases to over 50 GPa for $Fe^{3+}$ at both the A- and B-site. The transition pressure $P_{tr}$ of A-site $Fe^{3+}$ increases with $x$, for both the Al-free and the Al-bearing Pv; while B-site $Fe^{3+}$ is always in LS, which is also the case in pPv. The dependence of $P_{tr}$ on $x$ reflects the importance of Fe-M (M=Fe or Al) interactions. In pPv, $Fe^{3+}$ in the A-site is always in LS at above 100 GPa, except for $(Mg_{1-x}Fe_x)(Si_{1-x}Al_x)O_3$ with $x$=17%, where two pairs of A-B sites are occupied by one Fe-Fe and one Al-Al pair and the $Fe^{3+}$ at the A-site is in HS. Note that, although using the GGA or including the Hubbard U term would increase our predicted spin transition pressures (Hsu et al., 2012; Yu et al., 2012), at low iron concentrations, spin state has little effect on elastic properties (e.g. Stackhouse et al., 2007; Li et al., 2005b).

**Table 1**
**The stable spin configurations, denoted by the spin state of $Fe^{3+}$ at the A site - B site, of $Fe^{3+}$-$Fe^{3+}$ and $Fe^{3+}$-$Al^{3+}$ bearing Pv and pPv at T=0K and different pressures. No pressure correction is applied.**

| P[Gpa] | | Iron spin state | | | |
|---|---|---|---|---|---|
| | Pv | $(Mg_{1-x}Fe_x)(Si_{1-x}Fe_x)O_3$ | | $(Mg_{1-x}Fe_x)(Si_{1-x}Al_x)O_3$ | |
| | | x=6.25% | x=12.5% | x=6.25% | x=12.5% |
| 25 | | HS-LS, fm | HS-LS, fm | HS | (1)HS-HS, afm |
| 50 | | LS-LS, afm | HS-LS, fm | HS | (1)HS-HS, afm |
| 75 | | LS-LS, afm | LS-LS, afm | LS | (2)HS-LS, fm |
| 100 | | LS-LS, afm | LS-LS, afm | LS | (2)LS-LS, afm |
| 125 | | LS-LS, afm | LS-LS, afm | LS | (2)LS-LS, afm |
| | pPv | | | | |
| | | x=8.3% | x=16.7% | x=8.3% | x=16.7% |
| 100 | | LS-LS, fm | LS-LS, fm | LS | (2)HS-LS, fm |
| 125 | | LS-LS, fm | LS-LS, fm | LS | (2)HS-LS, fm |
| 150 | | LS-LS, fm | LS-LS, fm | LS | (2)HS-LS, fm |

(1) two Fe-Al pairs; (2) one Fe-Fe and one Al-Al pair.

By comparing enthalpies, we determine the Pv→pPv transition pressure for $MgSiO_3$ to be 90.9 GPa (or 94.9 GPa including our pressure correction to LDA), in agreement with previous DFT-LDA calculations (Oganov and Ono, 2004) as well as experiments (Shim, 2008). Comparing the enthalpies of $Fe^{3+}$-bearing Pv and pPv, we further found that the Pv→pPv transition pressure decreases with iron content $x$, reaching 80.6 GPa when $x$=8%; when Al exists, the transition pressure is less sensitive to $x$---93 GPa for $x$=8% and 87 GPa when $x$=12%. This also implies that the existence of Al increases the $Fe^{3+}$-bearing Pv→pPv transition pressure, which is consistent with the role of Al in $MgSiO_3$ (Zhang and Oganov, 2006a).

*3.2. Temperature and pressure dependence of elastic properties*

We first compare the elastic constants $C_{ij}$ of $MgSiO_3$ Pv and pPv from this work with previous calculations (Fig. 1). Our results reconfirm the overall consistency between DFT-MD and QHA and the disagreement at high temperatures, especially for $C_{11}$, $C_{13}$, and $C_{33}$ in pPv, as was noted previously (Stackhouse and Brodholt, 2007). More benchmarking comparisons, as well as complete lists on the elastic, seismic, and thermodynamic properties are given in the supplementary material.

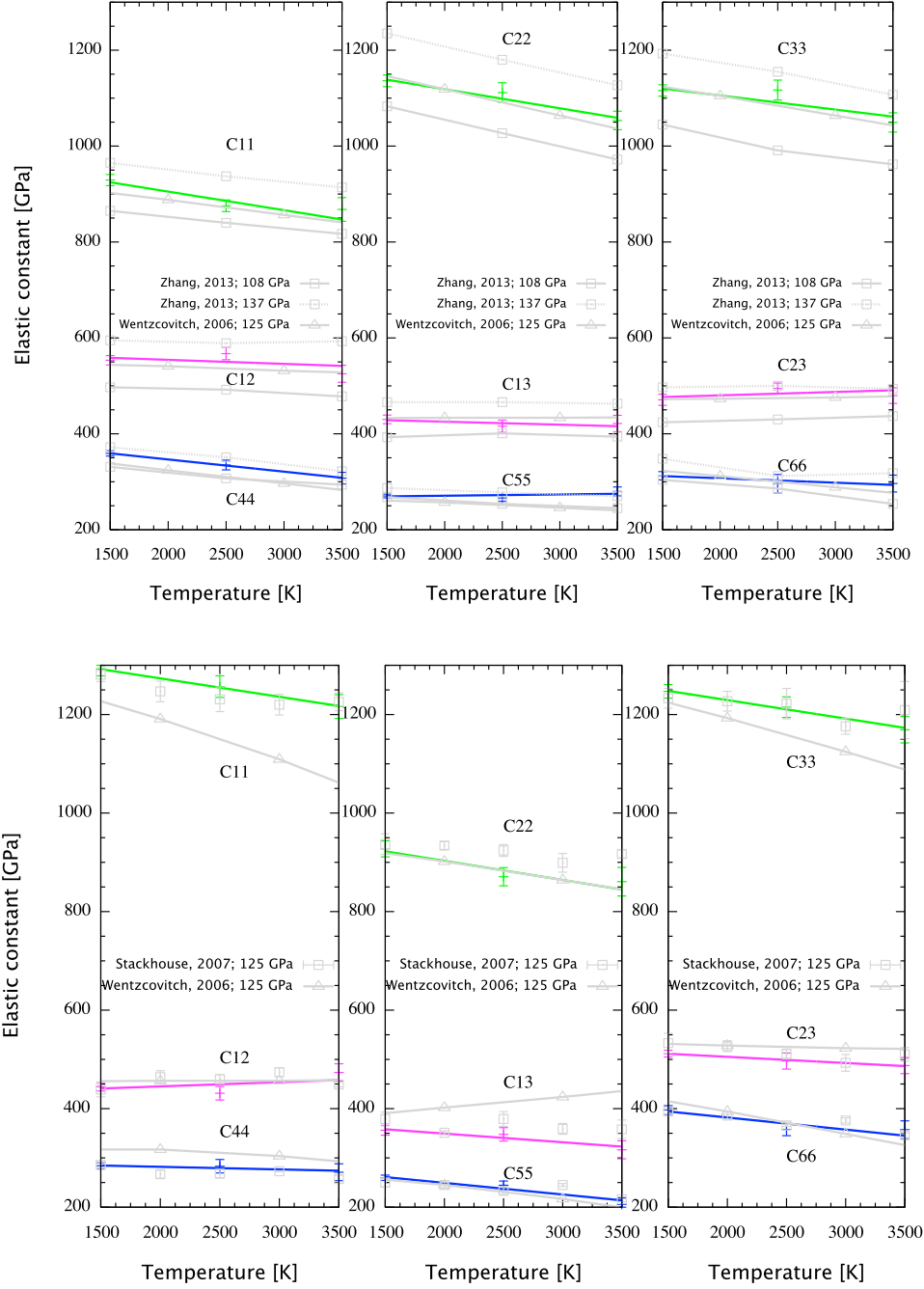

**Fig. 1 (color online). The elastic constants of MgSiO$_3$ Pv (top) and pPv (bottom) at 129 GPa in comparison with previous studies based on DFT-MD (squares) or QHA (triangles).**

Because the elastic properties depend on multiple variables ($T$, $P$, and $x$), we use the following equation to fit the computed elastic constants and the sound velocities

$$Q(T,P,x) = A_0(x)TP + A_1(x)T + A_2(x)P + A_3(x), \qquad (2)$$

where $Q$ is the property being fit, and each parameters $A_i(x)$ is linearly fit to $x$ by $k_i x + b_i$. Complete lists of the fit parameters can be found in the supplementary material.

Our results show that the elastic constants increase with pressure and mostly decrease with temperature, on the order of 1 for $dC_{ij}/dP$ compared to about $10^{-2}$ GPa/K for $dC_{ij}/dT$. $C_{12}$, $C_{13}$, and $C_{66}$ remain approximately constant when Fe or Al contents increases in Pv, while the other six values decrease. In contrast, for pPv, $C_{22}$, $C_{12}$ and $C_{13}$ increase with the Fe or Al contents. The effect of Fe is akin to that of Al and is usually accompanied by slightly higher $C_{ij}$ for Pv but lower $C_{ij}$ for pPv. The magnitude of the compositional derivatives depends on the specific P, T condition.

Analysis of the bulk and shear moduli ($K^S$ and $G$) as a function of P and T, using elastic constants fitted by Eq. (2), shows that $G$ uniformly decreases with $x$ for both Pv and pPv, but $K^S$ does not follow the same trend (Fig. 2). Both moduli are less sensitive to T than to P, similar to the dependence of $C_{ij}$, and their pressure derivatives are generally less dependent on $x$ at lower T, especially for Pv. The trend for ferrous iron is relatively simple---both moduli decrease with temperature, and the pressure derivatives are relatively more stable than the ferric iron bearing systems. This is understandable considering its more simple substitution mechanism and spin state, and fewer data at varied P and T.

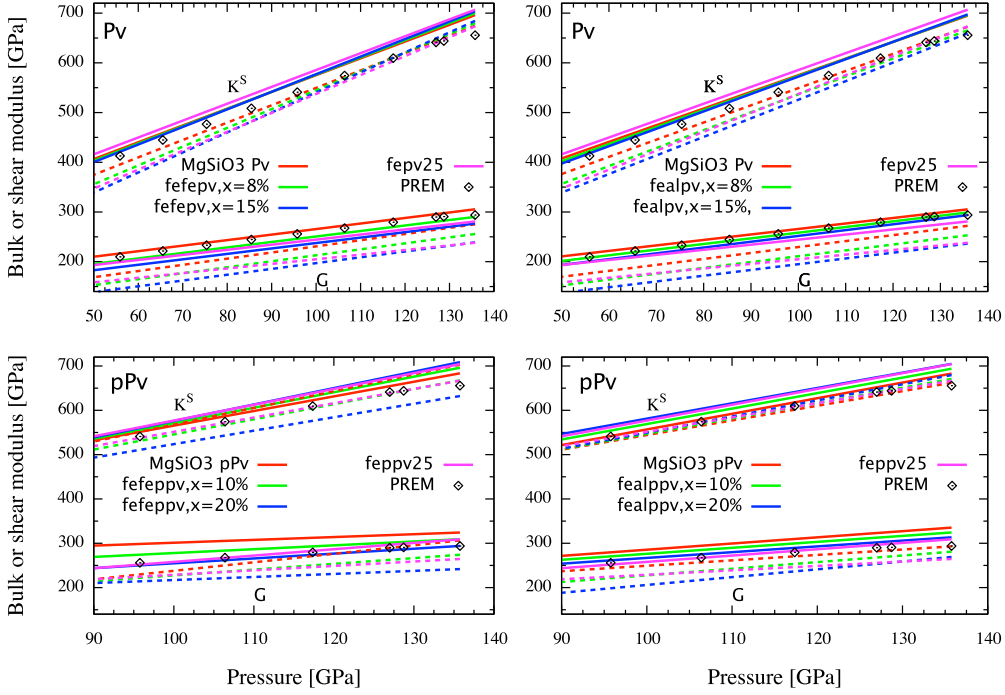

**Fig. 2 (color online). Bulk and shear moduli ($K^S$ and $G$) of pure and $Fe^{3+}$-$Fe^{3+}$ (fefe, left) or $Fe^{3+}$-$Al^{3+}$ (feal, right) bearing $MgSiO_3$ Pv and pPv at 2000 K (solid lines) and 4000 K (dashed lines), in comparison with that of $Mg_{0.75}Fe_{0.25}SiO_3$ (fepv25 and feppv25) and PREM as functions of pressure. Red, green, and blue meshes correspond to $x$=0, 8%, and 15% for Pv and 0, 10%, and 20% for pPv, respectively. In each panel, the top lines correspond to $K^S$, the lower ones correspond to $G$.**

### 3.3. Temperature, pressure, and composition dependence of sound velocities and density

The compressional ($V_p$) and shear ($V_s$) wave velocities are fitted as functions of P, T and $x$ using Eq. (2). The parameters are summarized in Table 2. In this way we can study how the velocities depend on the Fe or Al content $x$. As is shown in Fig. 3, the sound velocities steadily decrease with Fe or Al content, while the effect of Fe is larger, for both Pv and pPv. The pPv phase has higher velocities than Pv. At 110 GPa, the differences in $V_p$ and $V_s$ are 0.8% and 3.2%, respectively; the values further increase to 0.9% and 3.9% when $x$=5%, and to 1.1% and 4.5% when $x$=10%, for $MgSiO_3$ with $Fe^{3+}$-$Fe^{3+}$ pairs; with $Fe^{3+}$-$Al^{3+}$ pairs, the changes in $V_p$ are similar but that in $V_s$ are smaller by ~0.5%. The replacement of 25%

Mg by ferrous iron lowers the sound velocities by similar amounts to the effect of Fe-Al pairs on Pv and pPv.

Table 2
Fit parameters for the sound velocities of ferric iron bearing Pv and pPv. Vp, Vs, and Vb denote the compressional, shear, and bulk velocities, respectively.

|    | $A_0$-k | $A_0$-b | $A_1$-k | $A_1$-b | $A_2$-k | $A_2$-b | $A_3$-k | $A_3$-b |
|----|---------|---------|---------|---------|---------|---------|---------|---------|
| $(Mg_{1-x}Fe_x)(Si_{1-x}Fe_x)O_3$ Pv | | | | | | | | |
| Vp | 8.08E-06 | 1.84E-06 | -1.07E-02 | 2.00E-02 | -1.34E-03 | -3.42E-04 | -3.056 | 11.831 |
| Vs | -2.18E-06 | 1.76E-06 | 9.11E-03 | 5.18E-03 | -2.29E-04 | -3.55E-04 | -4.536 | 7.070 |
| Vb | 1.28E-05 | 6.62E-07 | -2.48E-02 | 2.10E-02 | -1.57E-03 | -1.06E-04 | 0.400 | 8.616 |
| $(Mg_{1-x}Fe_x)(Si_{1-x}Al_x)O_3$ Pv | | | | | | | | |
| Vp | 4.67E-06 | 1.58E-06 | 1.79E-03 | 2.01E-02 | -1.64E-03 | -3.26E-04 | -0.833 | 11.857 |
| Vs | -3.66E-06 | 1.56E-06 | 1.67E-02 | 5.24E-03 | -6.59E-04 | -3.44E-04 | -2.130 | 7.091 |
| Vb | 7.05E-06 | 5.55E-07 | -1.03E-02 | 2.09E-02 | -1.30E-03 | -1.01E-04 | 0.578 | 8.647 |
| $(Mg_{1-x}Fe_x)(Si_{1-x}Fe_x)O_3$ pPv | | | | | | | | |
| Vp | -1.24E-05 | 1.84E-06 | 7.98E-03 | 1.89E-02 | 6.50E-04 | -3.55E-04 | -4.145 | 12.086 |
| Vs | -3.50E-07 | 9.73E-07 | -2.70E-02 | 8.11E-03 | -4.81E-04 | -2.76E-04 | 0.213 | 7.006 |
| Vb | -1.50E-05 | 1.34E-06 | 3.34E-02 | 1.69E-02 | 1.10E-03 | -1.79E-04 | -5.011 | 8.968 |
| $(Mg_{1-x}Fe_x)(Si_{1-x}Al_x)O_3$ Pv | | | | | | | | |
| Vp | 9.90E-06 | 7.81E-07 | -1.98E-02 | 2.13E-02 | -1.62E-03 | -2.36E-04 | 0.788 | 11.802 |
| Vs | 1.82E-05 | 2.62E-09 | -3.48E-02 | 1.01E-02 | -2.84E-03 | -1.57E-04 | 3.225 | 6.752 |
| Vb | -4.43E-06 | 9.55E-07 | 5.98E-03 | 1.80E-02 | 5.30E-04 | -1.48E-04 | -1.742 | 8.852 |

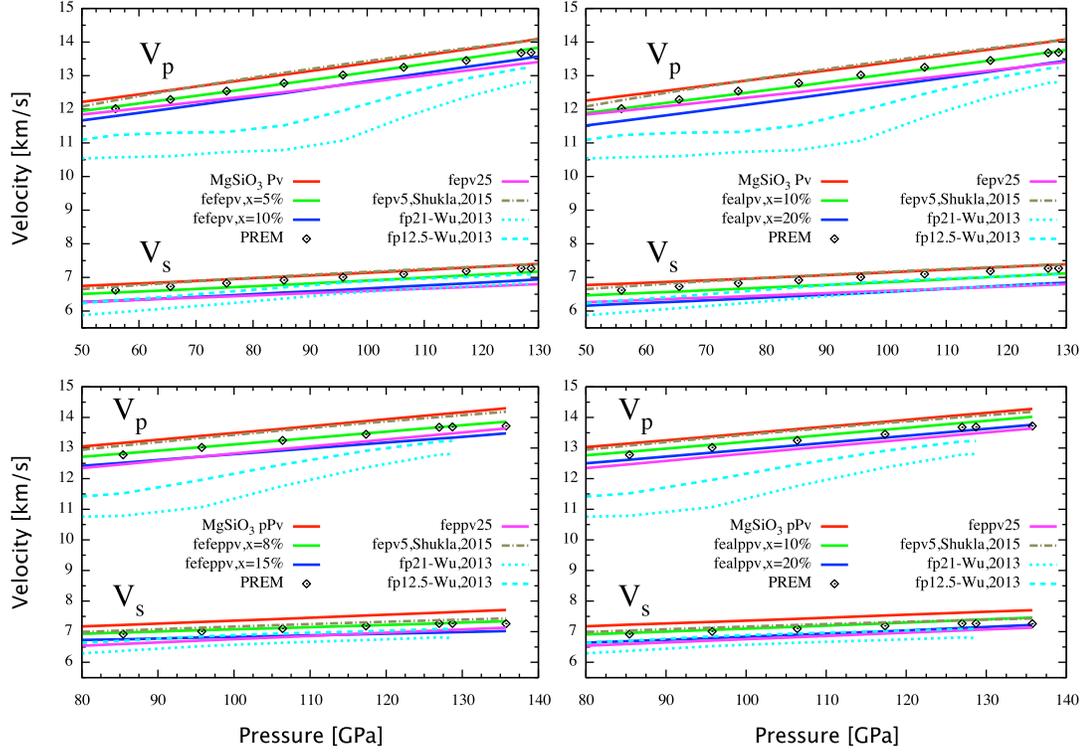

Fig. 3 (color online). The sound velocities of pure and $Fe^{3+}$-$Fe^{3+}$ (fefe, left) or $Fe^{3+}$-$Al^{3+}$ (feal, right) bearing $MgSiO_3$ in the Pv (top) and pPv (bottom) phases in comparison with that of $Mg_{0.75}Fe_{0.25}SiO_3$ (fepv25 and feppv25), $Mg_{0.95}Fe_{0.05}SiO_3$ [fepv5, from Shukla et al. (2015)], ferropericlase [fp21 and fp12.5, from Wu et al. (2013)], and the PREM along the geotherm by Brown and Shankland (1981).

We investigate the relation between density and P, T and $x$ by starting from the 3$^{rd}$-order Birch-Murnaghan equation of state (EoS). The EoS has three ambient parameters, $\rho_0$, $K_0$, and $K_0'$, and is an isothermal equation that has been widely used for minerals at Earth-interior conditions. Here, we assume linear dependence of $\rho_0$ and $K_0$, and independence of $K_0'$, on T, and fit P as a function of $\rho$ and T for each iron or aluminum content $x$. Finally, we fit the density at each P, T condition linearly to $x$, for systems with $Fe^{2+}$, $Fe^{3+}$-$Fe^{3+}$, and $Fe^{3+}$-$Al^{3+}$, respectively. We plot density along a geotherm (Brown and Shankland, 1981) and compare with ferropericlase and the preliminary reference Earth model (PREM) (Dziewonski and Anderson, 1981), as shown in Fig. 4. Our results show that, at 110 GPa, the density of the pPv phase is higher than that of Pv by 1.5%, 1.5%, 1.4%, and 1.3% for pure $MgSiO_3$, $Mg_{0.75}Fe_{0.25}SiO_3$, $Fe^{3+}$ or $Al^{3+}$ content $x$=6.25%, and $x$=12.5%, respectively.

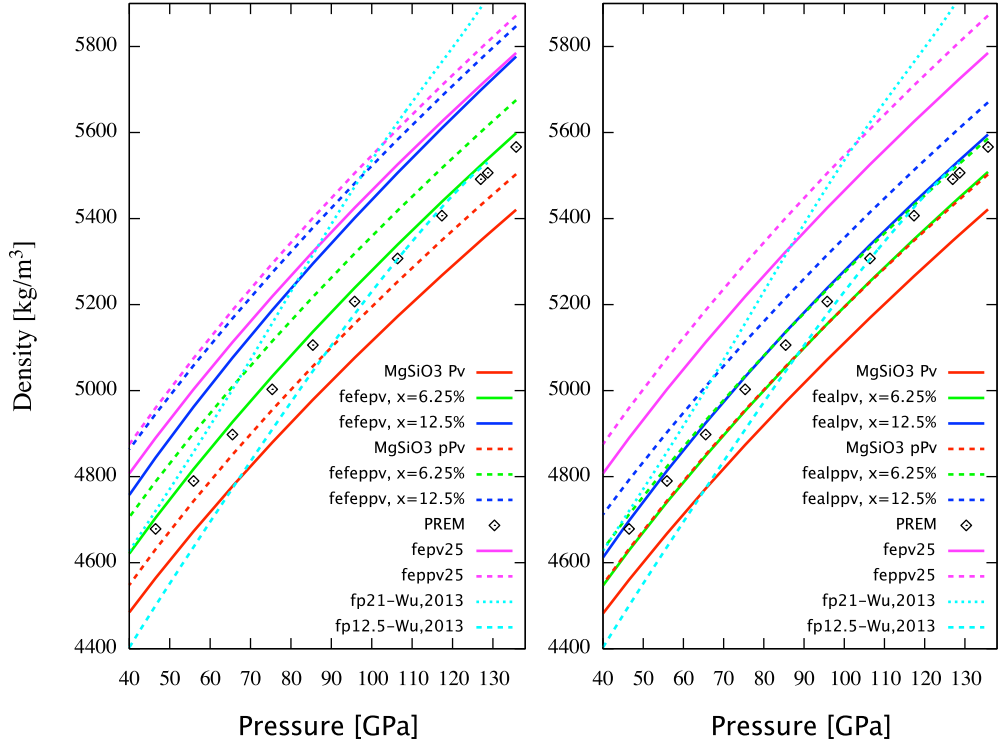

**Fig. 4 (color online).** The density profile of pure and $Fe^{3+}$- $Fe^{3+}$ (fefe, left) or $Fe^{3+}$-$Al^{3+}$ (feal, right) bearing $MgSiO_3$ in the Pv and pPv phases, in comparison with that of $Mg_{0.75}Fe_{0.25}SiO_3$ (fepv25 and feppv25), ferropericlase [fp21 or fp12.5, from Wu et al. (2013)] and PREM along the geotherm (Brown and Shankland, 1981).

*3.4. Elastic anisotropy*

The shear wave anisotropy $AV_S = \left|\frac{V_{S1}-V_{S2}}{(V_{S1}+V_{S2})/2}\right|$ of Fe/Al-bearing Pv and pPv are studied as functions of *P* and *T* (Walker and Wookey, 2012) for waves traveling along the three axes. As expected from the structural differences between Pv and pPv, we found much higher $AV_S$ in pPv than Pv (~10-30% vs ~10%). In Pv, shear waves travelling along [001] are slightly more anisotropic than those along [100] and [010], while in pPv it is the opposite. The dependence of anisotropy on P and T is weak in general, potentially explained by the stability of the structures with compression or heating in lower-mantle conditions. Our results (see supplementary Figs. S5-6) also suggest that the inclusion of Fe or Al increases the anisotropy of $MgSiO_3$ pPv but this is not the case for Pv.

## 4. Geophysical implications

### 4.1. Mineralogical composition of the lower mantle

The above calculations enable us to consider mineral assemblages and evaluate their density and sound velocities. By comparing these results with a seismic velocity model, such as the PREM, one can propose a mineralogical composition for the lower mantle, which is important for understanding the formation and evolution of the Earth and other planets, but impossible to directly determine.

To achieve this, we first examine a mixture of $Mg_{0.95}Fe_{0.05}SiO_3$:$Mg_{1-x}Fe_xSi_{1-x}Fe_xO_3$:$Mg_{1-x}Fe_xSi_{1-x}Al_xO_3$:$Fe_{0.125}Mg_{0.875}O$=34:10:43:13 (volume ratio), which corresponds to a mole ratio perovskite:ferropericlase=74:26, close to the pyrolitic composition that was proposed by Ringwood (1975) for the upper mantle, but is also widely assumed for the lower mantle. Here, the moduli and density values for $Mg_{0.95}Fe_{0.05}SiO_3$ and $Fe_{0.125}Mg_{0.875}O$ are based on Shukla et al. (2015) and Wu et al. (2013), respectively. Our calculation shows that, when choosing $x$=0.06 in $Mg_{1-x}Fe_xSi_{1-x}Al_xO_3$ and $x$=0.05 in $Mg_{1-x}Fe_xSi_{1-x}Fe_xO_3$, the velocities match PREM very well from 50 GPa to the top of the D″ (Fig. 5, left), and the agreement in density is also good (difference < 1%).

In such an assemblage, the number of cations per 12 O is Mg: 4.48, Si: 3.45, Fe: 0.37, Al: 0.11, which is remarkably consistent with that in pyrolite (Lee et al., 2004). Therefore, neglecting minor phases such as $CaSiO_3$, we can conclude that the major part of the Earth's mantle resembles the pyrolitic model. This is in contrast to the silicon-rich model for the Earth's lower mantle and implies that the extra silicon relative to the CI chondrite may reside in the D″ region or the core (Komabayashi et al., 2010).

At higher pressures when Pv can be less stable than pPv, we find that, with higher Al concentration in the pPv phase and less ferropericlase but with higher iron content, the velocities of the assemblage fit PREM values well in the high-pressure region (>105 GPa) (Fig. 5, right), although the density $\rho$ is slightly higher (by 2%). If such an assemblage exists, the per 12 O cation numbers (Mg: 4.16, Si: 3.22, Fe: 0.72 and Al: 0.32) indicate the same Mg/Si ratio, but more Fe and Al than the major part of the mantle above. This is also qualitatively consistent with the $Fe^{2+} \rightarrow Fe^{3+}$ transformation and varying $Fe^{3+}/\Sigma Fe$ with depth (Xu et al., 2015). However, there are shortcomings to this approach in the D″. First, the Brown and Shankland geotherm that we are using here neglects the steep thermal gradient expected near the core-mantle boundary (CMB). In the lowermost few hundred kilometers, the temperature could be underestimated by a few hundred to 1500 K, which in turn means that the velocities and the density of the mineral assemblage are overestimated (by about 1% for $V_p$, 3% $V_s$, and 1% for $\rho$). Including the temperature effects could support an assemblage with the same composition throughout the lower mantle (see Fig. S9 in the supplementary material), which currently overestimates $V_p$, $V_s$, and $\rho$ by 1.5%, 2.3%, and 0.4%, respectively, in the pPv region. Second, the Voigt averaging scheme for pPv is also associated by an uncertainty of ~1% for $V_p$ and $V_s$ (see supplemental Table S1). Lastly, both velocity and density variations are significant in the lowermost mantle and could reflect significant thermal and compositional variations (see supplementary material).

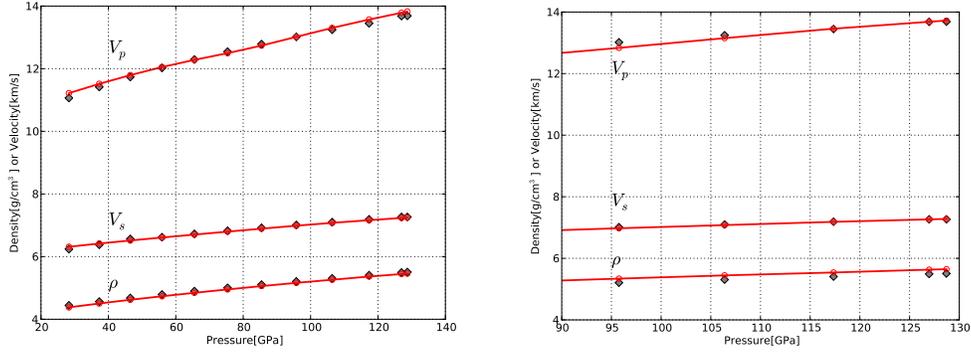

**Fig. 5 (color online). Comparing density and velocities of mineral assemblages with that of PREM (black diamond points) assuming the geotherm by Brown and Shankland (1981). Left, the assemblage consists of 34 vol.% $Mg_{0.95}Fe_{0.05}SiO_3$, 10 vol.% $Mg_{0.95}Fe_{0.05}Si_{0.95}Fe_{0.05}O_3$, and 43 vol.% $Mg_{0.94}Fe_{0.06}Si_{0.94}Al_{0.06}O_3$ in the Pv phase and 13 vol.% $Fe_{0.125}Mg_{0.875}O$; right: it consists of 26 vol.% $Mg_{0.95}Fe_{0.05}SiO_3$, 10 vol.% $Mg_{0.9}Fe_{0.1}Si_{0.9}Fe_{0.1}O_3$, and 53 vol.% $Mg_{0.85}Fe_{0.15}Si_{0.85}Al_{0.15}O_3$ in the pPv phase and 11 vol.% $Fe_{0.21}Mg_{0.79}O$. $Mg_{0.95}Fe_{0.05}SiO_3$ data are estimated with the BurnMan code (Cottaar et al., 2014b) by applying the thermodynamic model of Stixrude and Lithgow-Bertelloni (2011) on the calculation by Shukla, et al. (2015). For the pPv phase of $Mg_{0.95}Fe_{0.05}SiO_3$, $K^S$ is estimated to be the same as that of the Pv phase, $G$ and $\rho$ are estimated to be 5% and 1.5% larger than that of the Pv phase, respectively. Ferropericlase data are from Wu, et al. (2013).**

### *4.2. Seismic anisotropy*

While the lower mantle is thought to be mostly isotropic, seismic anisotropy is observed in D″. The main characteristic is that horizontally polarized waves travel faster than vertically polarized ones (or $\xi = V_{SH}^2/V_{SV}^2 > 1$) in regions beneath subducting slabs in regional studies as well as in global models. Observations of seismic anisotropy in D″ are often explained by crystal preferred orientation in post-perovskite (Cottaar et al. 2014a), but these studies have so far ignored major element composition. Based on the elastic constants in this study, we forward model the resulting texture and seismic anisotropy in perovskite and post-perovskite using the Viscoplastic Self-Consistent (VPSC) method (Lebensohn and Tomé, 1993) and deformation information from a typical tracer within a subducting slab (Cottaar et al., 2014a). While many slip systems are active within the model, we denote different experiments by the slip plane of their most active slip system. For perovskite this is (001), while for post-perovskite this is either (100), (010) or (001) (Cottaar et al., 2014a). For the perovskite model with dominant (001) slip we see an incompatible signature of $\xi<1$ across all compositions, pressures and temperatures, and we do not show these results here.

The resulting shear wave splitting and $\xi$ values for post-perovskite are shown in Fig. 6 at 100 GPa and 2500 K. Overall, we see that the transverse isotropy scales with the strength and orientation of splitting in the horizontal plane. The (100) and (001) cases show consistent signs in $\xi$ with major element composition: (100) produces a signature opposite from the one seen in seismology, while (001) is consistent with seismology. The resulting $\xi$ for the (010) case on the other hand is close to 1 and varies in amplitude; $\xi$ increases for $x$=8% compared to $x$=0%, but loses all amplitude when $x$=17%. Regardless of this trend, the amplitudes of $\xi$ for the (010) case are too weak to explain seismic observations. Generally, seismic observations will underestimate the actual amplitudes at depth, while our modeling will overestimate the amplitudes due to assuming pure dislocation creep as the deformation mechanism and by excluding the presence of periclase. This leaves the (001) case as the only compatible mechanism for seismic transverse isotropy, insensitive to the amount of Fe or Al included. None of the models predict strong azimuthal anisotropy, i.e. splitting through the vertical axis.

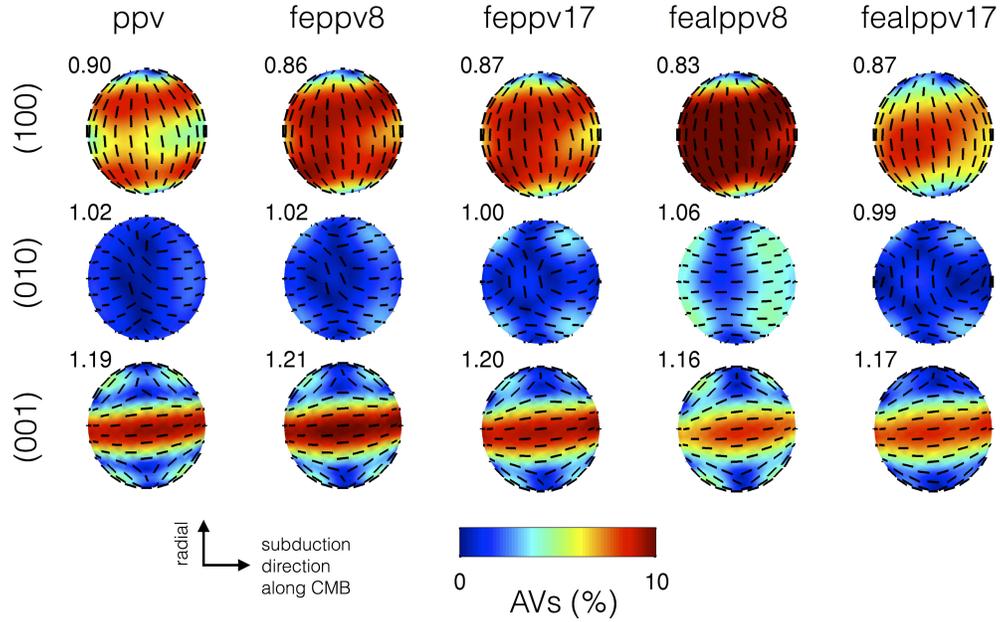

**Fig. 6 (color online).** Shear wave splitting strength (AVs, in color) and fast axis direction (black bar) shown as a function of seismic wave propagation direction on a sphere for the five different compositional model and three different sets of slips systems. The horizontal plane is parallel to the CMB with the direction of subduction to the right. The value for transverse isotropy, $\xi = V_{SH}^2/V_{SV}^2$, is given in the upper left corner of each subplot. Figure is made with Matlab Seismic Anisotropy Toolkit (Walker and Wookey, 2012) ppv: pure $MgSiO_3$ pPv; feppv8 or feppv17: $Fe^{3+}$-$Fe^{3+}$ bearing $MgSiO_3$ with $x=8\%$ or $17\%$; fealppv8 or fealppv17: $Fe^{3+}$-$Al^{3+}$ bearing $MgSiO_3$ with $x=8\%$ or $17\%$.

## 5. Conclusion

The thermoelasticity of pure, as well as Fe- and Al-bearing $MgSiO_3$ bridgmanite and post-perovskite have been calculated at a variety of P, T conditions and Fe or Al compositions. Our results show that the presence of Fe, especially $Fe^{3+}$, and Al results in complicated characteristics in the elastic and seismic properties of these minerals, depending on the their content and distribution, the phase of the mineral, and the P, T conditions. Our results, when combined with values previously calculated for (Fe,Mg)O, allow us to calculate the seismic properties of mineral assemblages and thus explore the probable composition of the lower mantle. We show that the average lower mantle seismic velocities and density can be fit with a pyrolitic composition. We also study the transverse anisotropy of these minerals in a subducting slab near the CMB, and confirm that the pPv phase with predominant (001) slip plane matches the seismic anisotropy in the D″ region independent of the amount of Fe or Al. These results provide a useful resource that can be used to investigate the effect of impurities on the properties of mineral assemblages in the deep mantle.

**Acknowledgement**

The calculations are performed using NCAR and NERSC supercomputing facilities. The authors appreciate Ian Rose, Rudy Wenk, Raymond Jeanloz, B. K. Godwal and Barbara Romanowicz for discussion and thank Mingming Li and Allen McNamara for providing the deformation information within a subducting slab. S. Zhang and B. Militzer are supported by NSF. S. Stackhouse and T. Liu were supported, in part, by NERC grant number NE/K006290/1. S. Cottaar is funded by the Drapers' Company Research Fellowship from Pembroke College, Cambridge, UK. We also acknowledge the use of high performance computing provided by Advanced Research Computing at the University of Leeds.

**Appendix A. Supplementary material**

Supplementary material related to this article has been attached.